\documentclass[twocolumn,prb,10pt,aps,longbibliography,superscriptaddress]{revtex4-2}

\usepackage{graphicx}
\usepackage{amsmath}
\usepackage{amssymb}
\usepackage{amsfonts}
\usepackage{dsfont}
\usepackage{dcolumn}
\usepackage{bbm}
\usepackage{bm}
\usepackage{mathtools}
\usepackage{csquotes}
\usepackage{braket}
\usepackage{physics}
\usepackage{siunitx}
\usepackage[unicode]{hyperref}
\usepackage[nameinlink]{cleveref}
\Crefname{section}{Sec.}{Secs.}
\Crefname{equation}{Eq.}{Eqs.}
\Crefname{figure}{Fig.}{Figs.}
\Crefname{tabular}{Tab.}{Tabs.}

\usepackage{bbold} 
\usepackage{nicefrac}
\usepackage{tikz}
\usetikzlibrary{arrows.meta}
\usepackage{xcolor}
\usepackage[makeroom]{cancel}

\newcommand{\bs}{\begin{subequations}}
\newcommand{\es}{\end{subequations}}
\newcommand{\be}{\begin{equation}}
\newcommand{\ee}{\end{equation}}
\newcommand{\vk}{{\vec k}}

\newcommand{\rr}{{\vec r}}

\definecolor{darkgreen}{rgb}{0.0, 0.6, 0.0}

\NewDocumentCommand\abos{ s g }{ \ensuremath{ \IfBooleanTF#1 { a^{\dagger}_{#2}} { a^{\phantom{\dagger}}_{#2} } } }
\NewDocumentCommand\bbos{ s g }{ \ensuremath{ \IfBooleanTF#1 { b^{\dagger}_{#2}    } { b^{\phantom{\dagger}}_{#2} } } }
\NewDocumentCommand\alp{ s g }{ \ensuremath{ \IfBooleanTF#1 { \alpha^{\dagger}_{#2}    } { \alpha^{\phantom{\dagger}}_{#2} } } }
\NewDocumentCommand\bet{ s g }{ \ensuremath{ \IfBooleanTF#1 { \beta^{\dagger}_{#2}    }{ \beta^{\phantom{\dagger}}_{#2} } }} 

\begin{document}

\title{Quantitative description of long-range order in the anisotropic spin-1/2 Heisenberg antiferromagnet on the square lattice}

\author{Nils Caci}
\email{nils.caci@lkb.ens.fr}
\affiliation{Laboratoire Kastler Brossel, Coll\`ege de France, CNRS, \'Ecole Normale Sup\'erieure - Universit\'e PSL, Sorbonne Universit\'e, 75005 Paris, France}
\affiliation{Institute for Theoretical Solid State Physics, RWTH Aachen University, Otto-Blumenthal-Str. 26, 52074 Aachen, Germany}

\author{Dag-Bj\"orn Hering}
\email{dag.hering@tu-dortmund.de}
\affiliation{Condensed Matter Theory, 
TU Dortmund University, Otto-Hahn-Stra\ss{}e 4, 44221 Dortmund, Germany}

\author{Matthias R.\ Walther}
\email{matthias.walther@fau.de}
\affiliation{Department of Physics, Friedrich-Alexander-Universit\"{a}t 
Erlangen-N\"urnberg (FAU), Staudtstra\ss{}e 7, 91058 Erlangen, Germany}

\author{Kai P.\ Schmidt }
\email{kai.phillip.schmidt@fau.de}
\affiliation{Department of Physics, Friedrich-Alexander-Universit\"{a}t 
Erlangen-N\"urnberg (FAU), Staudtstra\ss{}e 7, 91058 Erlangen, Germany}

\author{Stefan Wessel}
\email{wessel@physik.rwth-aachen.de}
\affiliation{Institute for Theoretical Solid State Physics, RWTH Aachen University, Otto-Blumenthal-Str. 26, 52074 Aachen, Germany}

\author{G\"otz S.\ Uhrig}
\email{goetz.uhrig@tu-dortmund.de}
\affiliation{Condensed Matter Theory, 
TU Dortmund University, Otto-Hahn-Stra\ss{}e 4, 44221 Dortmund, Germany}

\date{\textrm{\today}}

\begin{abstract}
  The quantitative description of long-range order remains a challenge in quantum many-body
  physics. We provide zero-temperature results from two complementary methods for the ground-state energy per site, the sublattice magnetization, the
  spin gap, and the transverse spin correlation length for the spin-1/2 anisotropic quantum Heisenberg antiferromagnet on the square lattice. 
  On the one hand, we use exact, large-scale quantum Monte Carlo (QMC) simulations. 
	On the other hand, we use the semi-analytic approach based on continuous similarity transformations in terms of elementary magnon excitations. 
  Our findings confirm the applicability and quantitative validity of
  both approaches along the full parameter axis from the Ising point to the symmetry-restoring phase transition at the Heisenberg point and further provide quantitative reference results in the thermodynamic limit.
In addition, we analytically derive the relation between the dispersion and the correlation length at zero temperature in arbitrary dimension, and discuss improved second-moment QMC estimators.
\end{abstract}

\maketitle

\section{Introduction}
\label{s:introduction}

The collective behavior of matter is one of the most important topics of modern
science. Understanding it is crucial in basic research as it holds the key to a
variety of correlated many-body states - realized for example in topologically
ordered quantum phases such as spin liquids and fractional quantum Hall liquids
or superconductors. At the same time, it is clear that this understanding forms
the basis for many technological applications such as magnetic data storage
\cite{chapp07} and spintronics \cite{baltz18,gomon17} which define the modern
era. In particular, it is decisive to gain a systematic and quantitative
understanding of collective phenomena.

Quantum magnetism has turned out to serve as a very fruitful playground to
study collective quantum phenomena. Indeed, frustrated magnets are known to
host a variety of exotic quantum phases such as spin-liquid ground states with
topological quantum order. Unfrustrated quantum antiferromagnets with
long-range magnetic N\'eel order play an important role for the physics of the
high-temperature superconductors \cite{bedno86}, because the undoped parent
compounds represent two-dimensional antiferromagnetic Heisenberg magnets with
$S=1/2$ on the square lattice \cite{manou91}. Furthermore, spintronics is a
huge field of applications in which the manipulation of magnetism is crucial
\cite{chapp07,gomon17,baltz18} requiring a fundamental understanding of the
underlying mechanisms. 

It is very common to describe the magnetic excitations of quantum
antiferromagnets by essentially interaction-free magnons \cite{auerb94}. By
``essentially free of interactions'' we mean that some static renormalization
of the dispersions is accounted for, but the magnon-magnon scattering processes
are rarely considered quantitatively. Yet it turned out that the interaction of
the magnons is very important to understand certain dips (dubbed `roton dips')
in the dispersion \cite{powal15,verre18b} and the distribution of spectral
weight in the continua \cite{powal18}. This has been shown by establishing a
continuous basis change, namely a continuous similarity transformation (CST),
such that the number of magnons in the target basis is conserved, at least to
very good approximation \cite{powal15,powal18,walth23}. This facilitates the
interpretation of the results greatly, for instance the dispersion can be read
off immediately. Continua at zero temperature formed from two or three
elementary excitations \cite{schmi22b} only require to solve a two- or
three-particle problem avoiding complicated diagrammatic techniques. In this
fashion, also bound states of two \cite{knett01b,windt01,notbo07,tseng23} or
three particles \cite{schmi22b} were identified. 

To support these calculations and to assess their validity, we turn to static
quantities such as the ground-state energy per site $e_0$, the sublattice
magnetization $m^z$, the spin gap $\Delta$, and the transverse spin correlation
length $\xi$ and compare theoretical results for them stemming from two very
different approaches. 

One is the above-mentioned CST which maps the original
model expressed by magnon creation and annihilation operators to an effective
model in magnon operators, but with conservation of the number of magnons. The
CST consists in setting up a general flow equation in all couplings and solving
this flow equation. The first step is analytical, the second numerical
eventually providing the coupling constants of a magnonic model in second
quantization. Thus, we classify this approach as semi-analytical.

The second approach are quantum Monte Carlo (QMC) simulations using the
stochastic series expansion (SSE) algorithm with directed-loops
\cite{Sandvik1991,Sandvik1999,Syljuasen2002,Alet2005}, which is a well
established numerically exact method for unbiased, large-scale studies of quantum
magnets.

Our work demonstrates the applicability and quantitative validity of CST and
QMC along the full parameter line from the Ising to the Heisenberg model. In
particular, we capture the phase transition from the gapped ordered phase at
any finite anisotropy to the gapless Heisenberg point with full SU(2) symmetry.
Our results confirm that associated critical exponents at this
symmetry-restoring phase transition are given exactly by the ones from
spin-wave theory.

The article is set up as follows. In the following section \ref{s:MaM} we
briefly introduce the model and describe the two employed complementary methods
concisely with an emphasis on how the particular quantities are
computed. The results are presented and discussed in Sect.\ \ref{s:results}.
The summary \ref{s:conclusions} concludes the article.

\section{Model and Methods}
\label{s:MaM}

\subsection{Model}
\label{ss:model}

We consider a Heisenberg quantum antiferromagnet with $S=1/2$ and 
easy-axis anisotropy defined by the anisotropy parameter 
$\lambda=J_{xy}/J_z\in [0,1]$ so that the Hamiltonian takes the form
\begin{equation}
      H= J \sum_{\langle i,j\rangle} \left[\frac{\lambda}{2}\left(S^+_iS^-_j + S^-_iS^+_j\right) + S^z_i S^z_j  \right] 
			\label{eq:model}
\end{equation}
in terms of the spin ladder operators. The model is bipartite and thus unfrustrated.
It breaks the $\mathds{Z}_2$ symmetry of spin flips $S^z_i \to -S^z_i$ by displaying
long-range alternating magnetic order with 
a finite sublattice magnetization $m$ at zero temperature. We assume $m$ to be positive
for simplicity, but its negative value is physically equivalent. At the spin isotropic
point $\lambda=1$ the continuous SU(2) symmetry is restored so that
the elementary excitations, the magnons, become Goldstone bosons
with vanishing spin gap $\Delta=0$ \cite{auerb94}. For any
value $0\le \lambda < 1$ a finite energy gap $\Delta>0$ is present, which takes the value $2J$ in the
Ising limit $\lambda=0$. No rigorously exact results are known for this model for $\lambda>0$. But there
is a wealth of evidence for its characteristic behavior \cite{manou91,auerb94,sandv97b}.
Since the singular behaviour at the Heisenberg point originate from Goldstone excitations and not by 
critical fluctuations, the associated exponents are expected to correspond exactly to the spin-wave values \cite{singh89a}. 
This has been confirmed by high-order series expansions about the Ising limit as well as CST for the closing of the
one-magnon gap $\Delta$ \cite{singh89a,dusue10,walth23} with an exponent $1/2$.
		
\subsection{Methods}
\label{ss:methods}

Next, we briefly sketch the two approaches used.

\subsubsection{Quantum Monte Carlo simulations}
\label{sss:qmc}

The starting point of the SSE QMC method is a high-temperature expansion of the quantum partition function

\begin{equation}
  Z = \mathrm{Tr} e^{-\beta H} = \sum \limits_{\left\{ \ket{\alpha}\right\}}\sum \limits_{n=0}^{\infty} \frac{\beta^n}{n!} \bra{\alpha} (-H)^n \ket{\alpha}\,.
  \label{eq:SSE_Z}
\end{equation}
Here, $\{\ket{\alpha}\}$ denotes an orthonormal basis of the full Hilbert space
and $n$ the expansion order of the exponential term. Further, $H$ is decomposed
into a sum of local bond operators $H = - \sum_{b,t} H_{b,t}$, characterized by a bond $b$ and an operator type $t$. These bond operators are chosen such that,
given the basis $\{\ket{\alpha}\}$, they are non-branching, i.e. $H_{b,t}
\ket{\alpha} \propto \ket{\alpha^\prime}, \forall
\ket{\alpha},\ket{\alpha^\prime} \in \{\ket{\alpha}\}$. 
The standard approach for the basis, which we also use here, is a product state
of local $\ket{S_i^z}$ eigenstates, i.e. $\ket{\alpha}=\bigotimes_{i=1}^{N}
\ket{S_i^z}$. Here, a suitable bond decomposition is given by the operators
\begin{equation}
  H_{b,1} = -J S_i^z S_j^z \, ,
  \label{eq:SSE_Hd_diag}
\end{equation}
that are diagonal in $\ket{\alpha}$ as well as the offdiagonal bond operators
\begin{equation}
  H_{b,2} = -\frac{J\lambda }{2} \left( S^+_i S^-_j + S^-_i S^+_j \right)\,  .
  \label{eq:SSE_Hb_offdiag}
\end{equation}

The different products of bond operators contributing to $Z$ can be encoded into
an ordered operator sequence $S_n=\left[(t_1,b_1),\dots,(t_n,b_n)\right]$, thus
yielding for the quantum partition function
\begin{equation}
  Z = \sum \limits_{\left\{ \ket{\alpha}\right\}}\sum \limits_{n=0}^{\infty} \sum \limits_{\left\{S_n\right\}} \frac{\beta^n}{n!} \bra{\alpha} H_{b_n,t_n}\cdots H_{b_2,t_2}H_{b_1,t_1}\ket{\alpha} \, .
  \label{eq:SSE_Z_Sn}
\end{equation}

The length of $S_n$ corresponds to the expansion order and thus fluctuates. For
numerical simulations, however, it is more convenient to truncate the sum to a
maximal expansion order $M$. Since the average expansion order is (sharply)
centered around $\langle n \rangle \propto N \, \beta$, this cutoff does not
introduce any systematic error in practice as long as we always ensure $M\gg
\langle n\rangle$. A fixed length operator string $S_M$ can be obtained by
padding $S_n$ with unity operators $H_{0,0}$, which finally yields for the
quantum partition function
\begin{equation}
  Z = \sum \limits_{\left\{ \ket{\alpha}, S_M\right\}} \frac{\beta^n (M-n)!}{M!} \bra{\alpha} \prod \limits_{p=1}^M H_{b_p,t_p}\ket{\alpha} \, ,
  \label{eq:SSE_Z_fixedM}
\end{equation}
where an additional combinatorial factor $\binom{M}{n}^{-1}$ has to be
introduced to account for the possible insertion positions of the unity
operators. Here, the expansion order $n$ corresponds to the number of non-unity operators in $S_M$.
For a bipartitite lattice, such as the square lattice considered here, all finite contributions to $Z$ in Eq.~\ref{eq:SSE_Z_fixedM} can be rendered positive, upon adding an appropriate  constant $C$ to the diagonal $H_{b,1}\rightarrow H_{b,1}+C$~\cite{Sandvik1991}.
To efficiently sample the configuration space $\{\ket{\alpha},S_M\}$
by means of Markov Chain Monte Carlo, a tandem of two updates is usually
carried out. In a first local update step \cite{Sandvik1991,Sandvik1999},
diagonal operators are inserted or removed in $S_M$, effectively sampling the
expansion order $n$. The next update is the directed-loop update
\cite{Syljuasen2002,Alet2005}. Here, by carrying out a succession of local bond
operator updates, a global and efficient update to sample different operator
types, i.e. both diagonal and offdiagonal, in $S_M$ as well as $\ket{\alpha}$
is obtained. For more details we refer the reader to
Refs.~\cite{Syljuasen2002,Sandvik2010}.

When approaching the Ising limit $\lambda \to 0$, we would ideally like our
updating procedure to reduce to a classical (cluster) update on the spins in
$\ket{\alpha}$. While this is indeed the case for the word-line based loop
algorithm (cf. Ref.~\cite{Evertz2003} for a review), which reduces in the Ising
limit to a classical Swendsen-Wang cluster update \cite{Swendsen1987} for
$\beta\to\infty$ \cite{Kawashima1995}, this is not the case for the
directed-loop update in the SSE picture for the following reasons. Flipping a
spin in $\ket{\alpha}$ here corresponds to flipping the spin on all operators
in $S_M$ that are connected to that site. For the local bond operator updates
during the directed-loop update, there is in the general case $\lambda \neq 1$
always the possibility to backtrack the loop propagation (this is also referred
to as a \enquote{bounce}).  As the probability for these bounces becomes large
for $\lambda\to0$, this process becomes very inefficient for $\beta\to\infty$,
since the number of operators connected to a site is proportional to $\beta$
\cite{Syljuasen2002}. A simple trick to overcome this issue is to perform a
unitary rotation in spin space, resulting in the Hamiltonian
\begin{equation}
  \tilde{H}= J \sum_{\langle i,j\rangle} \left[ S_i^x S_j^x + \lambda \left(S_i^y S_j^y +S_i^z S_j^z \right)\right] \, .
  \label{eq:H_spinrot}
\end{equation}

Rotating the Hamiltonian in spin space such that the largest matrix elements
are associated with off-diagonal bond operators more generally seems to help
with sampling issues that arise in the directed-loop update, as recently
reported in spin-1 systems with single-ion anisotropies \cite{Caci2021} or in
spin-1/2 $J Q_2$ models \cite{Liu2024}.

Thus, we always carry out simulations for the rotated Hamiltonian $\tilde{H}$.
In the following, we briefly discuss how we measure the relevant observables,
where $\langle \cdot \rangle_H$ ($\langle \cdot \rangle_{\tilde{H}}$)  refers
to thermal expectations with respect to $H$ ($\tilde{H}$). We further note,
that we always simulate quadratic lattice geometries with $N=L^2$ sites and
periodic boundary conditions.

In terms of the SSE, the energy is related to the average expansion order and
given by $E=-\langle n \rangle_H/\beta=-\langle n \rangle_{\tilde{H}}/\beta$.
To obtain an estimate of the ground-state energy $e_0$ per spin in the thermodynamic
limit, we simulate $E/N$ for linear system lengths $L=\{48,64,72\}$ at inverse
temperature $\beta=L$ and linearly extrapolate the data, using the leading
finite size behavior $L^{-3}$.

While the spectrum of the Hamiltonian is not directly accessible within the SSE
approach, it is indirectly encoded in imaginary time correlation functions
\begin{equation}
  C(\tau) = \langle \hat{O}(\tau) \hat{O}\rangle = \langle e^{\tau H}\hat{O}e^{-\tau H}\hat{O}\rangle \, ,
  \label{eq:QMC_imag_corrfunc}
\end{equation}
where $\hat{O}$ is some operator and $\tau \in [0,\beta]$ denotes the imaginary
time. If $\hat{O}$ is chosen such that it excites the ground state, i.e.
$\bra{\mathrm{GS}}\hat{O}\ket{\mathrm{GS}}=0$ and
$\hat{O}\ket{\mathrm{GS}}\neq0$, the ground-state excitation gap $\Delta$ can
be efficiently estimated in the thermodynamic limit (TDL) based on a (convergent) sequence of moments
of $C(\tau)$ as introduced in Ref.~\cite{Todo2015}. Here an order $m$ gap estimator $\Delta_{m,\beta}$ is
constructed using the Fourier components of $C(\tau)$
\begin{equation}
  \tilde{C}(\omega_k) = \int \limits_0^\beta C(\tau) e^{i\tau \omega_k} \mathrm{d}\tau \, ,
  \label{eq:QMC_Comega}
\end{equation}
where $w_k= 2\pi k /\beta\, , k\in \mathbb{Z}$ denote bosonic Matsubara
frequencies. The gap estimator $\Delta_{m,\beta}$ is then given as follows
\begin{equation}
  \Delta_{m,\beta} = \omega_1 \sqrt{-\sum \limits_{k=0}^{m}k^2 x_{m,k}\tilde{C}(\omega_k)/
    \sum \limits_{k=0}^{m}x_{m,k}\tilde{C}(\omega_k) } \, ,
  \label{eq:QMC_gap_estimator}
\end{equation}
where the numbers $x_{m,k}$ are defined by $x_{m,k} = 1/\prod_{j=0,j\neq k}^{m}(k+j)(k-j)$. Here, the important property 
\begin{equation}
  \lim \limits_{m\to \infty}\lim \limits_{\beta \to \infty}
  \Delta_{m,\beta}=\lim \limits_{\beta \to \infty}\lim \limits_{m\to \infty}
  \Delta_{m,\beta}=\Delta
  \label{eq:QMC_gap_convergence}
\end{equation}
holds, where the limits are in particular interchangeable. For more details we
refer the reader to Ref.~\cite{Todo2015}. A suitable operator for the one-magnon
excitations is the staggered magnetization in the spin-$x$ direction $\hat{M}^x$
such that we consider $C(\tau) = \langle \hat{M}^x(\tau)\hat{M}^x \rangle_H =
\langle \hat{M}^z(\tau)\hat{M}^z \rangle_{\tilde{H}}$. For the gap estimator
$\Delta_{m,\beta}$, we measure $\tilde{C}(\omega_k)$ with respect to the
rotated system $\tilde{H}$. In practice, as outlined in Ref.~\cite{Todo2015}, a
sufficient choice for convergence of $\Delta_{m,\beta}$ is given by $m=5$ and
$\beta=2\pi/\tilde{\Delta}$, where $\tilde{\Delta}$ is an initial guess of
$\Delta$, for which we use the CST results. To eliminate finite-size effects,
we choose linear system lengths of $L\approx 2\beta$, including system
lengths up to $L=384$ near the isotropic Heisenberg point.

The staggered longitudinal magnetization $m^z$ can be obtained as $m^z = \sqrt{\langle
(\hat{M}^z/N)^2\rangle_H}=\sqrt{\langle (\hat{M}^x/N)^2\rangle_{\tilde{H}}}$ from the spin-spin
correlation function at maximal distance $\vec{r}_\mathrm{max}=(L/2,L/2)$ \cite{Sandvik1997,Reger1988,Sandvik2010},
\begin{equation}
  \langle (\hat{M}^x/N)^2 \rangle_{\tilde{H}} = 
  \begin{cases}
    \phantom{3}\langle S_{i}^x S_j^x \rangle_{\tilde{H}}(r_{ij}= \left|\vec{r}_\mathrm{max}\right|), &\text{if } \lambda < 1\\
    3\langle S_{i}^x S_j^x \rangle_{\tilde{H}}(r_{ij}=\left|\vec{r}_\mathrm{max}\right|), &\text{if } \lambda = 1 
  \end{cases},
  \label{eq:CSXSX}
\end{equation}
where $r_{ij}=\left|\vec{r}_{i}-\vec{r}_{j}\right|$ and the factor $3$ is
included for $\lambda=1$ to account for the SU(2) symmetry. The offdiagonal
spin-spin correlation function $\langle S_i^x S_j^x\rangle_{\tilde{H}}$ can be
efficiently measured during the directed-loop update \cite{Dorneich2001,
Alet2005}. To obtain an estimate of the ground-state magnetization $m^z$, we
extrapolate data for linear system sizes $L=\{48,64,72\}$ at inverse
temperatures $\beta=L$ to the TDL using the leading
finite-size scaling behavior $L^{-1}$ (here it is important to take the square
root after the extrapolation). Close to the isotropic Heisenberg point, the
finite-size effects are stronger, and we here linearly extrapolate data for
linear system sizes $L=\{96,128,140,160\}$ at inverse temperatures $\beta=L$.

The transverse correlation length $\xi^x$ quantifies the exponential asymptotic decay
of the spin-spin correlation function $\langle S_i^x S_j^x\rangle_{H}\propto
\mathrm{exp}(-r_{ij}/\xi^x)/r_{ij}$. It  is always finite for $\lambda<1$  and relates to
the quantum fluctuations that arise for $\lambda \neq 0$. 
In addition to 
the dominant exponential term an algebraic factor $1/r_{ij}$ appears 
for the two-dimensional system considered here. 
In App.~\ref{app::disp}, we provide a general derivation 
of the algebraic correction factor within a saddle point 
approximation for general dimensions, and identify its 
actual presence explicitly for the system under 
consideration from QMC simulations in Sec.~\ref{s:results}. 
One possible means of extracting the correlation length $\xi^x$ is thus to perform a fit of the numerical data to this asymptotic decay within appropriate ranges of the distance $r_{ij}$. 
Another  standard  approach to estimate $\xi^x$ is by means of the staggered spin structure factor
$S^x(\vec{q})$, defined by
\begin{align}
  S^x(\vec{q}) &= \frac{1}{N} \sum \limits_{i,j} (\pm1) e^{-i \vec{q}\cdot (\vec{r}_i-\vec{r}_j)} \langle S_i^x S_j^x \rangle_{H} \nonumber\\
  &= \frac{1}{N} \sum \limits_{i,j} (\pm 1)e^{-i \vec{q}\cdot (\vec{r}_i-\vec{r}_j)} \langle S_i^z S_j^z \rangle_{\tilde{H}} \, ,
  \label{eq:QMC_Sq_estimator}
\end{align}
where for the staggering sign \enquote{+1} is chosen if the sites $i$ and $j$
belong to the same sublattice and \enquote{-1} otherwise. In terms of this quantity, $\xi^x$ can be estimated by the
 second-moment estimator
\begin{equation}
  \xi^x = \sqrt{2} \frac{1 }{\left|\vec{q}_1\right|}\sqrt{\frac{S^x(\vec{0})}{S^x(\vec{q}_1)}-1} \, ,
  \label{eq:QMC_xi_estimator}
\end{equation}
where $\vec{q}=\vec{0}$ corresponds to the antiferromagnetic ordering and
$\vec{q}_1= \vec{e}_x2\pi/L$ is one of the reciprocal lattice vectors closest
to $\vec{0}$. The factor $\sqrt{2}$ is specific to the algebraic correction
term $1/r_{ij}$ in the spin-spin correlation function, cf.\ App.~\ref{app::est}
for other cases. For our analysis, we focus on the regime $\lambda \geq 0.8$,
where efficient simulations can be carried out without performing the spin rotation. 
To obtain better statistics, we thus measure $\langle S_i^x S_j^x \rangle_H$ 
during the directed-loop update. Further, we extrapolate the data to the TDL
using polynomials of degree 3 in $1/L$ for linear system sizes up to $L=100$ 
and at inverse temperature $\beta=1/L$.

 
\subsubsection{Continuous Similarity Transformations}
\label{sss:cst}

For the semi-analytic CST approach we rewrite the spin model \eqref{eq:model}
in terms of bosons according to Dyson and Maleev \cite{dyson56a,malee58}
which hides the manifest hermiticity of the Hamiltonian. But it ensures
that the bosonic Hamiltonian 
\begin{align}
  H &= J \sum_{i \in \Gamma_A, \delta}  \big[ - S^2 
	\nonumber\\
  &+ S \left( \abos*{i}\abos{i} +\bbos*{i+\delta}\bbos{i+\delta} + 
	\lambda \abos{i} \bbos{i + \delta} +  \lambda \abos*{i} \bbos*{i + \delta} \right)
	\\ \nonumber 
	&- { \abos*{i} \abos{i} \bbos*{i+\delta}  \bbos{i+\delta} }
      -\frac{\lambda}{2} { \abos*{i} \abos{i}  \abos{i} \bbos{i+\delta} }
   -\frac{\lambda}{2} { \abos*{i} \bbos*{i+\delta}  \bbos*{i+\delta} 
			\bbos{i+\delta} } \big],
\end{align}
comprises at maximum quartic terms in the bosonic creation and annihilation operators.
Here the sum runs over the sites of the $A$ sublattice $\Gamma_A$ where the $a$ bosons
are located. The $b$ bosons are located 
on the $B$ sublattice $\Gamma_B$ while the $\delta$ link nearest neighbors,
i.e., a site on $\Gamma_A$ to one of its adjacent sites on $\Gamma_B$.

The next step is to Fourier transform the bosonic operators to $\abos{\bm{k}}$ and
$\bbos{\bm{k}}$ and their hermitian conjugates, which yields the Hamilton operator
with bilinear and quartic bosonic terms in reciprocal space with wave vectors ${\bm k}$.
These wave vectors are chosen from the magnetic Brillouin zone because 
we distinguish between both sublattices.  Subsequently, we perform
a standard Bogoliubov transformation to operators $\alp{\bm k}$ and $\bet{\bm k}$
to eliminate the bilinear terms creating or annihilating pairs of bosons.
This is done self-consistently, i.e., the bilinear terms changing the number of bosons
vanish after normal-ordering all quartic terms. Normal-ordering is meant here relative
to the bosonic vacuum after the Bogoliubov transformation: all creation operators
are commuted to the left, all annihilation operators to the right.
Truncating the resulting Hamiltonian at the bilinear level provides the usual
mean-field Hamiltonian which neglects all interactions between the elementary
excitations, but comprises a static renormalization of the dispersion on
the level of a self-consistent Hartree-Fock theory.
	
We proceed by keeping all quartic terms which generally consist of three
classes of terms: (i) leaving the number of bosons invariant, (ii) changing this
number by $\pm2$, and (iii) changing this number by $\pm4$. Class (i) consists
of terms with two creation and two annihilation operators, class (ii) consists of terms with
three creation and one annihilation operator or vice-versa. Finally,
class (iii) consists of terms with four creation operators or four annihilation operators.

This quartic Hamiltonian is not manifestly hermitian because of the 
properties of the Dyson-Maleev representation. Thus, we apply a CST to it which is not manifestly unitary. 
The flow of coupling 	constants results from
\be
\label{eq:flow-eq}
\frac{d}{d\ell}H(\ell) = [ \eta(\ell),H(\ell)], 
\ee
where $\ell$ is the  continuous flow parameter running from $\ell=0$ where $H$ equals the
initial Hamiltonian to $\ell=\infty$ where one obtains an effective
Hamiltonian $H_\text{eff}= H(\infty)$ \cite{wegne94,kehre06}.
The initial Hamiltonian results from the self-consistently determined Bogoliubov
transformation while the final effective
Hamiltonian $H_\text{eff}$ does no longer
contain terms of classes (ii) and class (iii). In order to reach this nice property,
which facilitates the subsequent interpretation, we choose the particle
conserving generator $\eta_\text{pc}(\ell)$ \cite{knett00a,fisch10a} which comprises
all terms in the Hamiltonian at $\ell$ of classes (ii) and (iii). Those
terms which increase the number of bosons have exactly the same prefactor
as in $H(\ell)$ while those terms which decrease the number of bosons
have the opposite sign in the  prefactor of the same term in $H(\ell)$.

Assuming convergence of the flow implies $\eta(\infty)=0$ so that $H_\text{eff}$
\emph{conserves} the number of excitations, here magnons.
Then, the energy of single-magnon states can be read off directly
from the dispersion $\omega(\bm{k})$ without any further many-body corrections.
In order to be able to solve \eqref{eq:flow-eq} two further approximations
are necessary. First, we solve the problem for a finite cluster of $L^2$ sites
for various boundary conditions so that there is a finite number of points
in reciprocal space. Second, on the right hand side of \eqref{eq:flow-eq}
quartic terms are also commuted with quartic terms so that hexatic terms 
are generated. First, we normal-order them so that their constant, bilinear,
and quartic content is kept. But the genuine hexatic terms are not included 
because of their higher scaling dimension \cite{powal15,powal18,walth23}. 
This introduces a small truncation error for intermediate anisotropies.
But close to the isotropic point $\lambda=1$ we hardly detect
truncation errors \cite{walth23}.

The constant arising in $H_\text{eff}$ is the ground-state energy $E_0$.
The total sublattice magnetization $M^z$ in $z$-direction
 can be found as derivative of the ground-state energy
with respect to an alternating magnetic field $h_\text{alt}$
\be
\label{eq:magnetiz}
M^z = \frac{d}{dh_\text{alt}} E_0\Big|_{h_\text{alt}=0}
\ee
according to the Hellmann-Feynman theorem.
In practice, we compute the derivative by the ratio 
\mbox{$(E_0(h_\text{alt})-E_0(0))/h_\text{alt}$}
for small $h_\text{alt}\approx 10^{-7}J$.

The ground-state energy per site and the sublattice magnetization per site are denoted $e_0$ and $m^z$,
respectively. The magnon dispersion is denoted $\omega(\bm{k})$; its minimum defines the spin gap $\Delta$ 
which is located at $\vec{k}=(\pi,\pi)$. The correlation length is computed based on the full dispersion 
choosing a direction $\bm{n}=\bm{e}_x$ in  reciprocal space and solving
\be
\label{eq:correlation}
\omega(i\kappa\,\bm{n})=0 \quad \text{with} \quad \text{Re}\,\kappa=1/\xi.
\ee
according to a generalization of the results in Ref.\ \cite{okuni01} from one to two dimensions,
(see App.~\ref{app::disp} for the generalization to any spatial dimension including also leading power law corrections).
In the dispersions studied here, we never found a value of $\kappa$ with a finite imaginary part.
Finally, the results obtained for $L^2$ sites for various boundary conditions
are extrapolated to the thermodynamic limit $L\to\infty$. We use results 
for $L\in\{17,19,21\}$ and antiperiodic boundary conditions because odd linear 
lengths and these boundary conditions display
the smallest dependence on the finite size of the evaluated cluster. 
For the technical details, we refer to  Refs.\ \cite{powal15,powal18,walth23}.

\section{Results}
\label{s:results}

\begin{figure}[htb]
    \centering
   \includegraphics[width=\columnwidth]{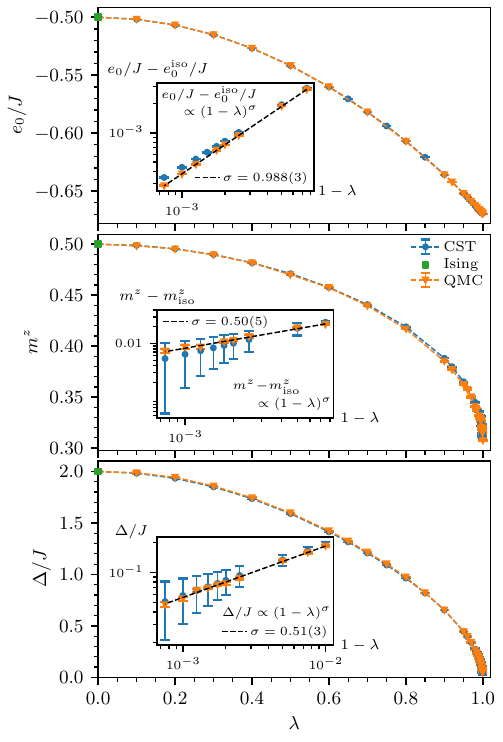}
    \caption{Comparison of ground-state data obtained by CST (circles) and QMC (triangles) both extrapolated
      to the TDL. The points at $\lambda=0$ (green marker)
      denote the exact result of the 2D Ising model. All QMC errors are
      statistical and smaller than the marker size. Upper panel:
      ground-state energy per site $e_0$. The relative statistical QMC errors
      are of order $O(10^{-6})$. The systematic error of the CST data is
      estimated to be $O(10^{-5})$. 
			The inset shows that the energy approaches its isotropic value 
			linearly agreeing with previous results \cite{hamer92,zheng05}.
			Middle panel: sublattice magnetization per site $m^z$. The relative statistical QMC errors are of
      order $O(10^{-3})$; the CST error is $O(10^{-4})$. In addition, we 
      presume that the CST has a small systematic error. 
			The inset shows that the sublattice magnetization 
			approaches its isotropic value like a square root agreeing with mean-field results 
			\cite{hamer92,zheng05}.
			Lower panel: spin gap $\Delta$ with  errors shown in the inset. The data provides 
			strong evidence for $\Delta \propto \sqrt{1-\lambda}$ as assumed previously \cite{hamer92,zheng05}.}
    \label{fig:vergleich}
\end{figure}

The key results of the ground-state properties of the anisotropic spin-1/2
Heisenberg AFM on the square lattice are collected in Fig.\
\ref{fig:vergleich}. The CST data for the ground-state energy, 
spin gap,  and the correlation length have been published in Ref.\ \onlinecite{walth23} already. 
Overall, the QMC and CST data agree extremely well.

In the ground-state energy per site $e_0$, the relative errors of the QMC data
are of order $O(10^{-6})$. The systematic error of the finite-size
extrapolation is negligible, i.e. much smaller than the statistical errors.
The error of the CST data is systematic since no stochastic aspects are involved.
Two main error sources can be identified: (i) the truncation of the flowing Hamiltonian
to quartic order and (ii) the finite-size effect. The finite-size effect
can be assessed by varying the extrapolations and the boundary conditions,
and we estimate it to be $O(10^{-5})$.
The effect of the truncation cannot be estimated intrinsically, but only
by comparison to results from other methods. We find that above $\lambda
\approx 0.4$, the CST data start to deviate from the QMC data. This deviation
appears to be systematic, since it does not change in sign and furthermore evolves
smoothly. The deviation becomes maximum around $\lambda \approx 0.9$, where 
it is of order $O(2\cdot 10^{-4})$. Still the agreement is very good.

Repeating the estimates for the sublattice magnetization per site $m^z$, we
arrive at relative statistical QMC errors of order $O(10^{-3})$. In terms of
the CST, the error of the finite-size effect and the effect of approximating a
derivative by a ratio in the CST data is estimated as $O(10^{-4})$. The deviation
between the data from QMC and CST becomes here maximal around $\lambda\approx 0.9$ as well
and is of the order of $O(5\cdot10^{-3})$, which is $\approx 1.3\%$. 
This deviation, similarly to the deviation observed for the energy, does
not change sign and evolves relatively smoothly. Thus, we presume that 
it is systematic  and can be
attributed to the CST truncation. Still, the agreement of the data from the
two distinct approaches over the whole parameter range is very good.

\begin{figure}[htb]
    \centering
    \includegraphics[scale=1]{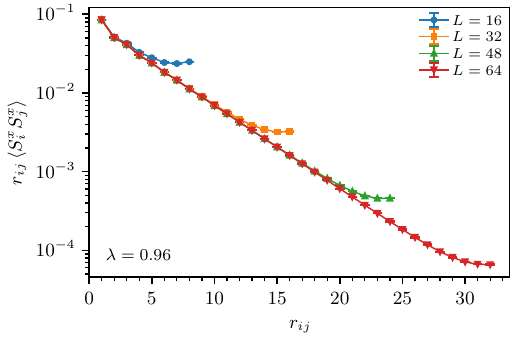}
    \caption{Rescaled transverse spin-spin correlations  $r_{ij} \langle S^x_i S^x_j\rangle$, obtained from QMC simulations at $\lambda=0.96$, as functions of the distance
    $r_{ij}$ for systems of different linear system sizes $L$, plotted in
  logarithmic form in order to probe the approach towards the anticipated
  asymptotic behavior.}
    \label{fig:corr}
\end{figure}

Next, we address the value of the spin gap $\Delta$. The relative
statistical errors of the QMC data are of order $O(10^{-3})$ and near the isotropic
Heisenberg point of order $O(10^{-2})$. The relative error estimate for the CST data
yields  $O(10^{-4})$ and lower for $\lambda<0.9$ and rises up to $0.66\%$ close to the isotropic point.
While there is a small systematic deviation of the CST data,
both approaches agree very well. We highlight that the
double-logarithmic plot in the inset of the lowest panel in Fig.\
\ref{fig:vergleich} displays an excellent agreement. Both data sets support the
conclusion that the spin gap vanishes in a square root fashion $\propto
\sqrt{1-\lambda}$, as reported in Refs.~\cite{hamer92,zheng05}. 
Similarly, the magnetization approaches its value for the isotropic Heisenberg model also following 
a square root law.

\begin{figure}[htb]
    \centering
    \includegraphics[width=\columnwidth]{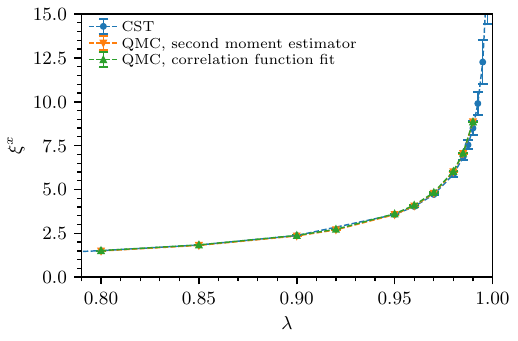}
    \caption{Ground-state transverse correlation length $\xi^x$ as function of
    $\lambda$ in the infinite size limit obtained by CST (circles) and QMC (triangles).}
    \label{fig:corr-length}
\end{figure}

Finally, we consider the correlation length $\xi$.  Since we are dealing with a
long-range ordered phase at all values of $\lambda$ in the ground state of the
Hamiltonian \eqref{eq:model} this leads to an infinite (longitudinal)
correlation length in the $S^z$ direction. But the magnons stand for
transversal fluctuations of this order parameter. 
Hence, we consider the transversal correlation length 
$\xi^x$ pertaining to the dominantly exponential decay of 
 $\langle S^x_i S^x_j\rangle \propto \exp(-r_{ij}/\xi^x)/r_{ij} $ for large values of
$ r_{ij}$. 

As shown in App.~\ref{app::disp} for this two-dimensional system, an additional
algebraic factor $1/r_{ij}$ appears in the asymptotic scaling. We can
demonstrate the above scaling behavior also based on QMC simulations. This is
illustrated for the case of $\lambda=0.96$ in Fig.~\ref{fig:corr}. As we
consider systems with periodic boundary conditions, the correlations $\langle
S^x_i S^x_j\rangle$ for a system size $L$ are shown for distances $r_{ij} \leq
L/2$. We find that the rescaled quantity $r_{ij} \langle S^x_i S^x_j\rangle$
approaches an exponential decay, as anticipated. By fitting the rescaled QMC
data for $L=64$ to an exponential decay within the regime of $r_{ij}$ between 5
and 20, we thus obtain a robust estimate for the correlation length $\xi^x$ in
the TDL. We also used the improved second-moment estimator in
Eq.~\ref{eq:QMC_xi_estimator} in order to obtain the correlation length $\xi^x$
from the structure factor which is extrapolated to the TDL including systems
up to $L=100$ with $\beta=L$. In CST, we describe the dispersion
$\omega(\bm{k})$ obtained numerically by a sum of cosine terms such that the
dispersion is exactly captured. Then, we determine the correlation length by
solving for the zeros in \eqref{eq:correlation}. Again, as shown in
Fig.~\ref{fig:corr-length}, the agreement is very good. 

Concerning the power laws upon approaching the isotropic Heisenberg point,
the CST and the QMC data sets strongly support the linear behavior of the 
ground-energy and square-root behavior for the sublattice magnetization,
the spin gap, and the correlation length. The latter two power laws are
not independent, but linked according to $\xi \propto 1/\Delta$. All these
power laws agree with the results of spin wave theory \cite{hamer92} which may
appear surprising. But we emphasize that approaching the isotropic point does
not represent a true quantum phase transition because the system stays in the 
same long-range ordered phase. The observed gap closure does not indicate 
a second order transition, but the restoration of the continuous symmetry
of spin rotation which in turn implies the occurrence of Goldstone bosons.
Since no critical quantum fluctuations appear the exponents remain the same as in
mean-field theory. Our findings are further in agreement with very recent DMRG
studies \cite{Kadosawa2024}, that we became aware of during the completion of our
manuscript.

\section{Conclusions}
\label{s:conclusions}

In this paper we examined the ground-state properties of the anisotropic
spin-1/2 Heisenberg AFM on the two-dimensional square lattice in the
thermodynamic limit. For this purpose we considered two approaches,
namely unbiased SSE QMC simulations as well as the semi-analytic CST method.

Based on our analysis of the ground-state energy $e_0$, the excitation gap
$\Delta$, the sublattice magnetization $m^z$, as well as the transverse
correlation length $\xi^x$, we report a very good
agreement of both approaches over the whole parameter range, from the Ising
limit $\lambda=0$ to the SU(2) symmetric isotropic Heisenberg point
$\lambda=1$. Our findings support the quantitative validity of both
approaches. In terms of the CST approach, this is particularly
interesting as the method may be applied to extract quantum-critical properties 
as well as dynamical correlation functions of frustrated systems such as the $J_1$-$J_2$ model 
or the Heisenberg model on the triangular lattice, where the statistical accuracy 
of QMC methods is severely limited, due to the sign problem.

Our data are available in \cite{Caci2024data}  and provide quantitative reference results in
the thermodynamic limit. We envision this to be of particular use for
bench-marking purposes of numerical approaches for strongly correlated quantum
spin systems.

\begin{acknowledgments} 
This work has been
financially supported by the Deutsche Forschungsgemeinschaft (DFG, German Research
Foundation) in project UH 90/14-1 (GSU) and SCHM 2511/13-1 (MRW/KPS) as well as
the RTG 1995 (NC/SW). NC further acknowledges support by the ANR through grant
  LODIS (ANR-21-CE30-0033) and thanks the IT Center at RWTH Aachen University
for access to computing time. 

KPS gratefully acknowledges the support by the Deutsche Forschungsgemeinschaft 
(DFG, German Research Foundation) -- Project-ID 429529648—TRR 306 \mbox{QuCoLiMa} 
(``Quantum Cooperativity of Light and Matter'') and the Munich Quantum Valley, 
which is supported by the Bavarian state government with funds from the 
Hightech Agenda Bayern Plus. MRW/KPS thankfully
acknowledge the scientific support and HPC resources provided by the Erlangen National
High Performance Computing Center (NHR@FAU) of the Friedrich-Alexander-Universit\"at
Erlangen-N\"urnberg (FAU).
\end{acknowledgments}

\appendix

\section{Correlation function from the one-particle dispersion}\label{app::disp}

This appendix contains a generalization of the results of Ref.~\cite{okuni01} from one 
to any spatial dimension concerning the computation of the correlation length based on 
the full one-particle dispersion.	
In addition, we also derive the leading power law corrections.

We consider a translationally invariant system on a lattice 
in which the elementary excitations are bosonic and display a gap.
Note that an ordered antiferromagnet on a bipartite lattice can be made translationally
invariant by a $180^\circ$ rotation of the spins on one sublattice so that it belongs
to the considered kind of systems.
The Hamiltonian reads
\be
H = \sum_{\vk\in\text{BZ}} \left[ A_\vk b^\dag_\vk b^{\phantom{\dag}}_\vk +\frac{1}{2} B_\vk (b^\dag_\vk b^\dag_{-\vk}+
b^{\phantom{\dag}}_\vk b^{\phantom{\dag}}_{-\vk})\right] + H_{\rm I},
\ee
where $b_\vk^{(\dag)}$ are the usual bosonic annihilation (creation) operators and $H_{\rm I}$ an interaction term
which needs not be specified further. The lattice constant is henceforth set to unity.
Neglecting the interaction, the dispersion reads
\be
\omega_\vk^2 = A_\vk^2- B_\vk^2 > 0
\ee
and the correlation $G_\rr := \langle b^\dag_j b^{\phantom{\dag}}_i\rangle$ with $\rr:=\rr_j-\rr_i$ is given 
at zero temperature by
\bs
\begin{align}
G_\rr &= \frac{1}{(2\pi)^d} \int_\text{BZ} \langle b_\vk^\dag b^{\phantom{\dag}}_\vk\rangle \exp({\rm i}\rr\cdot\vk) {\rm d}k^d
\\
&= \frac{1}{2(2\pi)^d} \int_\text{BZ} \left( \frac{A_\vk}{\omega_\vk}-1\right)\exp({\rm i}\rr\cdot\vk) {\rm d}k^d.
\label{eq:corr1}
\end{align}
\es
If we include interaction effects, the functions $A_\vk$ and $B_\vk$ will be modified and hence
$\omega_\vk$. For simplicity, we refrain from introducing new labels for these modified quantities.
In addition, Eq.\ \eqref{eq:corr1} does not hold anymore in a rigorous sense 
because of multi-boson contribution.
But multi-boson contributions form continua in the $(\omega,\vk)$ space which induce dependences in real space
which decay quicker than the contributions of the $\delta$-distributions resulting from the single-boson
states. However, the weights of the single-boson states are reduced due to hybridization with
multi-boson states. Introducing a weight factor $Z_\vk$ yields for the long-range part $G_\rr^\text{lo-ra}$ of the
correlation
\be
\label{eq:long-range-corr}
G_\rr^\text{lo-ra} = \frac{1}{2(2\pi)^d} \int_\text{BZ} Z_\vk \frac{A_\vk}{\omega_\vk}
\exp(i\rr\cdot\vk) {\rm d}k^d
\ee
where we also left out the constant background stemming from the summand $-1$ in \eqref{eq:corr1}.

From \eqref{eq:long-range-corr} one realizes that a significant contributions results from small
values of the dispersion. But this observation is not yet sufficient to find the correlation length.
For this, the saddle point approximation needs to be invoked. For clarity, we first discuss the
one-dimensional case, cf.~ Ref.\ \onlinecite{okuni01}.

\paragraph{One dimension}
The dispersion can be described by a sum of trigonometric functions so that it is analytic. Note that
the finite gap $0< \Delta \le \omega_k$ avoids the occurrence of singularities. The same holds true
for the numerator $Z_k A_k$. Then we can shift the integration from the real axis to a path $\gamma$ 
in the complex plane crossing a point where $\omega_{k_0}=0$
\be
\label{eq:long-range-corr2}
G_x^\text{lo-ra} = \frac{1}{4\pi} \int_\gamma Z_\vk \frac{A_\vk}{\omega_\vk}
\exp({\rm i} xk) {\rm d}k
\ee
In order to deal with the singularity in the above integrand, we substitute $k$ by $k(z)$ fulfilling
\be
\frac{{\rm d}k}{{\rm d}z}=\omega_k \quad \Leftrightarrow \frac{{\rm d}z}{{\rm d}k}=1/\omega_k .
\ee
We denote by $z_0$ the point in $\mathds{C}$ with $k(z_0)=k_0$. The chain rule implies
\be
k''=\frac{{\rm d}^2k}{{\rm d}z^2}= \frac{{\rm d}\omega_k}{{\rm d}k} \frac{{\rm d}k}{{\rm d}z} 
=\frac{{\rm d}\omega_k}{{\rm d}k} \omega_k = \frac{1}{2} \frac{{\rm d} (\omega^2_k)}{{\rm d}k}.
\ee
By $\widetilde\gamma$ we denote the contour of which the image is the original contour, i.e.,
$\gamma=k(\widetilde \gamma)$ with $k(z_1)=-\pi$ and $k(z_2)=\pi$. Then we can write
\be
\label{eq:long-range-corr-z}
G_x^\text{lo-ra} = \frac{1}{4\pi} \int_{\tilde\gamma} Z(z) A(z)
\exp({\rm i} xk(z)) {\rm d}z.
\ee
This integration can be evaluated for $x\to\infty$ by the saddle point approximation,
also known as method of the steepest descent, yielding
\be
\label{eq:saddle1}
G_x^\text{lo-ra} = \frac{1}{4\pi}  Z(z_0) A(z_0)
\exp(i xk_0) \sqrt{\frac{2\pi}{x|k''_0|} },
\ee
where $k''_0:=\frac{{\rm d}^2k}{{\rm d}z^2}\Big|_{z=z_0}$. With $1/\xi=\Im{k_0}$ we obtain
\be
\label{eq:corr-1D}
G_x^\text{lo-ra,1D} \propto \frac{\exp(-|x|/\xi)}{\sqrt{|x|}}.
\ee
For $x\to-\infty$, we repeat the above reasoning for $-k_0$ which also fullfills
$\omega_{-k_0}=0$ if the dispersion is even as is the case for systems with
inversion symmetry. 

We stress that the above derivation yields the power law correction $1/\sqrt{|x|}$ in addition
to the exponential decrease. Note that this power law differs from the classical
Ornstein-Zernicke law \cite{Cardy1996}.

\paragraph{Two dimensions}
In essence, we repeat the arguments of the one-dimensional case. But the integration
over the wave vectors is two-dimensional. We solve this issue by assuming rotational
invariance of the asymptotic behavior in space, i.e., only $r=|\rr|$ matters. 
Correspondingly, we assume that the integrand in \eqref{eq:long-range-corr} is 
rotationally invariant in reciprocal space. This is not completely true, but 
can be justified \emph{a posteriori}: we found that $\xi$ hardly depends on
the direction $(1,0)^\top$ or $(1,1)^\top$. The difference is only a few percent
in the range $\xi\approx 1$ and  vanishes exponentially for larger correlation lengths.

Under the assumption of rotational invariance Eq.\ \eqref{eq:long-range-corr}
becomes
\bs
\begin{align}
\label{eq:long-range-corr-2d}
G_\rr^\text{lo-ra} &\approx \frac{1}{8\pi^2} \int_0^\pi\! \int_0^{2\pi}\!\! k Z_k \frac{A_k}{\omega_k}
\exp({\rm i}rk\cos\varphi) {\rm d}\varphi {\rm d}k
\\
&= \frac{1}{4\pi} \int_0^\pi  k Z_k \frac{A_k}{\omega_k} J_0(rk) {\rm d}k. 
\end{align}
\es
The upper limit of the $k$-integration is an approximation; but one does not
need to bother because it does not enter the saddle point approximation.
The function $J_\nu(z)$ is the $\nu$-th Bessel function of the first kind.
Its asymptotic behavior for large arguments is 
\be
\label{eq:bessel-asymp}
J_\nu(z) \propto \Re \exp({\rm i}z-\pi(\nu+1/2)/2)/\sqrt{z}
\ee
so that we can
apply again the substitution and the saddle point approximation as in the
one dimensional case yielding finally
\be
\label{eq:corr-2D}
G_r^\text{lo-ra,2D} \propto \frac{\exp(-r/\xi)}{r}
\ee
with $1/\xi=\Im(k_0)$ from $\omega_{k_0}=0$ as before.
Note that the additional power law results from the asymptotic behavior 
of the Bessel function which in turn stems from the angular integration.
It reflects how large the contribution of $\exp({\rm i}rk)$ is. 

\paragraph{Three dimensions}
We repeat the arguments for the two-dimensional case. Assuming rotational invariance
as in two dimensions the integration over the solid angle yields 
\bs
\begin{align}
\frac{1}{4\pi}\int_\Omega {\rm d}\Omega \exp({\rm i}rk\cos(\vartheta))
&= \frac{1}{2} \int_{-1}^1 \exp({\rm ir}k u) {\rm d}u
\\
& = \frac{1}{rk} \Im \exp({\rm i}rk).
\end{align}
\es
As to be expected, the exponential dependence remains the same, but the power law
exponent is lowered by $1/2$. This leads us to 
\be
\label{eq:corr-3D}
G_r^\text{lo-ra,3D} \propto \frac{\exp(-r/\xi)}{r^{3/2}}
\ee
with $1/\xi=\Im(k_0)$ from $\omega_{k_0}=0$ as before.

\paragraph{Arbitrary dimension}
The obvious generalization of the above findings reads 
\be
\label{eq:corr-aD}
G_r^\text{lo-ra,aD} \propto \frac{\exp(-r/\xi)}{r^{d/2}}
\ee
in $d$ dimensions with $1/\xi=\Im(k_0)$ from $\omega_{k_0}=0$.
Note that $d-1$ factors of $\sqrt{r}$ in the denominator result
from a phase space argument or geometrical dilution. But one factor
$\sqrt{r}$ stems from the saddle point approximation, see the derivation
in one dimension.

Eq.\ \eqref{eq:corr-aD} can be derived as before. First, we integrate in \eqref{eq:long-range-corr}
over all angles at fixed modulus $q$.  We choose $\rr$ to point along
$\vec e_1$ so that $\rr\cdot\vk = rk_1$ and the angular integration yields
\bs
\begin{align}
f(r,q)  &= \int_\text{BZ} e^{{\rm i}\rr\cdot\vk}\delta(k-q ) {\rm d}k^d
\\
&= \iint_{-q}^q  e^{{\rm i} r k_1} \delta(k-q) {\rm d}k_\perp^{d-1} {\rm d}k_1
\\
&= \int_{-q}^q \int_0^q  e^{{\rm i} r k_1}\delta(k-q) 
\rho_{d-1}(k_\perp) {\rm d}k_\perp {\rm d}k_1
\end{align}
\es
where we denoted $k$ for $\sqrt{k_1^2+k_\perp^2}$ 
and used $\vec k_\perp$ for the wave vector perpendicular to $\vec e_1$ and the
density of the modulus $k_\perp$ is given in dimension $d$ by $\rho_d(k_\perp) \propto k_\perp^{d-1}$.
The integration over $k_\perp$ is carried out with the help of the $\delta$-distribution
yielding
\bs
\begin{align}
f(r,q)  &= q \int_{-q}^q  e^{{\rm i} r k_1} \frac{\rho_{d-1}(\sqrt{q^2-k_1^2}) }{\sqrt{q^2-k_1^2}} {\rm d}k_1
\\
&\propto q \int_{-q}^q e^{{\rm i} r k_1} (q^2-k_1^2)^{(d-3)/2} {\rm d}k_1
\\
& =q^{d-1} \int_{-1}^1 e^{{\rm i} r qx} (1-x^2)^{(d-3)/2} {\rm d}x
\\
&= q^{\nicefrac{d}{2}} \sqrt{\pi} \left(\frac{2}{r} \right)^{\nicefrac{d}{2} -1} \Gamma(\nicefrac{(d-1)}{2}) 
J_{\nicefrac{d}{2}-1}(rq)
\end{align}
\es
where we substituted $k_1=q x$. In the final integration over $q$, we use again the asymptotics 
\eqref{eq:bessel-asymp} and employ the saddle point approximation 
to obtain \eqref{eq:corr-aD} stated above.

Eq.\ \eqref{eq:corr-aD} constitutes the powerful generalization of the one-dimensional
result in Ref.\ \cite{okuni01} to arbitrary dimension including the leading multiplicative 
power law corrections.

We finally note that the above result in Eq.~(\ref{eq:corr-aD}) results from replacing in the  Ornstein-Zernike law for the correlation function~\cite{Cardy1996} in $d$ spatial dimensions, 
$G^\mathrm{OZ}(r)\propto \mathrm{exp}(-r/\xi_x)/r^{(d-1)/2}$, the dimension $d$
by $d+1$, corresponding to the usual quantum-to-classical mapping.

\section{QMC estimator for the correlation length}\label{app::est}

In this appendix, we discuss the QMC estimator for the  correlation length in different spatial dimensions and for different correlation function asymptotics. 
In the case of the Ornstein-Zernike form of the correlation function~\cite{Cardy1996} in $d$ spatial dimensions, 
$\langle S_i^x S_j^x\rangle \propto \mathrm{exp}(-r_{ij}/\xi^x)/r_{ij}^{(d-1)/2}$, the improved second-moment estimator reads
\begin{equation}
  \xi^x = \sqrt{\frac{8d}{(1+d)(3+d)}}\frac{1}{\left|\vec{q}_1\right|}\sqrt{\frac{S^x(\vec{0})}{S^x(\vec{q}_1)}-1} \, ,
  \label{eqB:QMC_xi_estimator}
\end{equation}
as discussed in Ref.~\cite{Sandvik2010} 
(note that there is an apparent typo in  Eq.~(71) of Ref.~\cite{Sandvik2010}). The numerical prefactor in the above estimator is of order one and thus often ignored when extracting  correlation lengths based on this estimator.  
Indeed, for $d=1$ and $d=3$ the prefactor simplifies to $1$ exactly,  while for $d=2$ it reads $\sqrt{16/15}$, explicitly. For other algebraic factors in the correlation function asymptotics, the numerical prefactor however differs. 
In particular, for the asymptotic behavior $\langle S_i^x S_j^x\rangle \propto \mathrm{exp}(-r_{ij}/\xi^x)/r_{ij}^{d/2}$ derived in the preceding appendix we similarly obtain, upon analytically performing the Fourier transformation in the continuum limit, the corresponding prefactors $\sqrt{8/3}$, $\sqrt{2}$, and $\sqrt{4/3}$ for $d=1,2,3$, respectively.

\section{Quantitative comparison of CST and QMC results}

\begin{figure}[htb]
    \centering
   \includegraphics[width=\columnwidth]{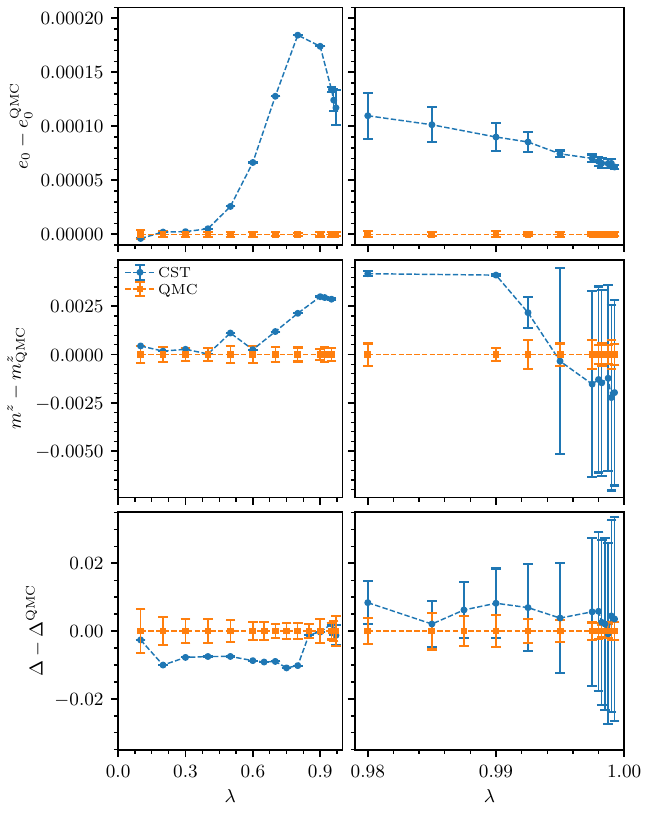}
   \caption{Quantitative comparison of the CST and QMC results in the TDL of the ground-state energy 
	$e_0$ (top panels), 
	the ground-state magnetization $m^z$ (middle panels),
	and the spin gap $\Delta$ (bottom panels) as a function of $\lambda$. }
    \label{fig:quant_vergleich}
\end{figure}

Here, we show a quantitative comparison of the CST and QMC results 
in terms of the statistical and systematic errors. In Fig.\ \ref{fig:quant_vergleich}
the deviations of the CST data relative to the QMC data is plotted for
the ground-state energy per site $e_0$, the magnetization per site $m^z$, and the energy
gap $\Delta$. In energy and magnetization the deviation is remarkably small 
for small values of $\lambda$ growing for large $\lambda$. We attribute this to
the truncation of the tracked terms on quartic level corresponding to scaling dimension $2$.
Hence, it is a systematic deviation which we had observed before in the binding energies
of two-magnon bound states \cite{walth23}. We do not have a complete understanding
of the deviations of the spin gap which is rather constant. Still, we presume a systematic
origin linked to the truncation in the CST approach. Remarkably, the deviations 
decrease in all three quantities upon approaching the isotropic point. 
We interpret this as a justification of the truncation according to scaling dimension
which allows us to find the relevant effective model close to the isotropic point
almost quantitatively.

\begin{figure}[htb]
    \centering
   \includegraphics[width=\columnwidth]{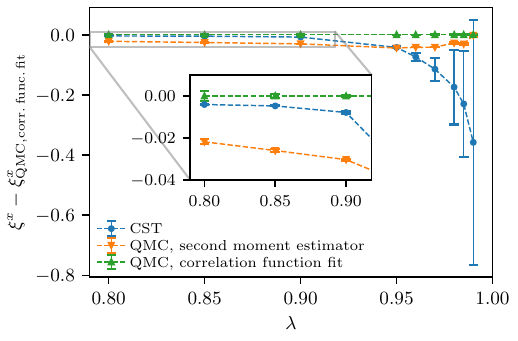}
   \caption{Quantitative comparison of CST (circles) and QMC (triangles) 
	results of the ground-state transverse correlation length $\xi^x$ 
      extrapolated to the thermodynamic limit as a function of $\lambda$.}
    \label{fig:quant_vergleich_xi}
\end{figure}

Finally, Fig.\ \ref{fig:quant_vergleich_xi} displays the deviation of 
three ways to access the transversal correlation length. The two QMC estimators
agree well below the one percent level. Such minor deviations may result from
uncertainties in performing the actual finite-size fitting process. The CST results
acquire large errors close to the isotropic point because the correlation
length depends inverse proportionally on the spin gap so that tiny inaccuracies
in the latter induce large inaccuracies in the correlation length.

\bibliography{liter11.bib}

\end{document}